# Analysis of electron localization in a coupled quantum dot structure via variational approach: numerical results and application in quantum-dot cellular automata

George A.P. Thé[1], Rubens V. Ramos[2], Sérgio A. de Carvalho[2]

[1]*Dipartimento di Elettronica, Politecnico di Torino, Corso Duca degli Abruzzi 24, 10129 Torino, Italia*

[2]*Group of Quantum Information, Department of Teleinformatics Engineering, Federal University of Ceara Campus do Pici, 725, C.P.6007,60755-640, Fortaleza-Ceará,Brasil*

**Abstract**

Analysis of quantum dot structures is a current topic with important applications in solid-state digital logic design, quantum information technology and quantum optics. In this work, we show a variational formulation for the solution of the effective two-level approach of the analysis of electron localization in two coupled quantum dots. Numerical results are presented as well the use of electron localization in the realization of a configurable logic circuit using quantum-dot cellular automata.

**1. Introduction**

Coherent destruction of tunneling (CDT) has been the subject of many papers in the last years and it occurs when the tunneling of a particle under the influence of a suitable external AC driving field is suppressed [1]. This suppression is directly related to the parameters of the driving field as, for instance, frequency and the envelope function of the electric field considered.

Firstly recognized in quantum well systems and superlattices, CDT has also been observed in quantum dot systems. In these nanostructures, the concentration of carriers is much lower than that present in quantum wells and superlattices and, because of that, the Coulomb interaction must be taken into account in the study of the dynamical localization and tunneling of electrons. Quantum dots are of great interest to the fields of quantum communication, in which are used to produce single-photon and entangled-photons sources

[2-5], to implement passive quantum key encoding [6] and to perform quantum teleportation [6-8], and in quantum computation too [9-11].

In this work we consider a system formed by two quantum dots and one electron under the influence of an external AC electric field. Instead of using the Hamiltonian of the system to derive the equations which govern the electron localization, we propose to use the Lagrangian formalism to find a most suitable set of equations, in the sense it requires lower computational efforts for a numerical solution than that obtained from the Hamiltonian. Once the localization of the electron was studied according to the characteristics of the electrical field applied, we consider the control of a quantum cellular automata cell by the control of the electron localization in coupled dots structure.

This paper is organized as follows: in the second section we show the theoretical model of the system and the set of nonlinear coupled differential equations obtained from the Hamiltonian and the Schrödinger equation. After we propose a Lagrangian to the system and, using the Euler-Lagrange equations, we find the set of coupled equations that is solved by the suggestion of an ansatz. The numerical simulations performed and their results are shown in Section 3. In Section 4, the control of a quantum cellular automata cell is considered and, at last, the conclusions are presented in Section 5.

## 2. Description of the dynamic of coupled quantum dots using Hamiltonian and Lagrangian

The coupled quantum dots are considered as a two-level quantum system and its quantum state, expanded in terms of the eigenstates of each isolated quantum dot, is

$$|\psi\rangle = a_R |R\rangle + a_L |L\rangle \tag{1}$$

where $|R\rangle$ and $|L\rangle$ are, respectively, the eigenstates of the $Q_R$ (right) and $Q_L$ (left) quantum dots. The Hamiltonian of the unperturbed two coupled quantum dots system is [12]

$$H_0(t) = \hbar k \left(|L\rangle\langle R| + |R\rangle\langle L|\right) + \varepsilon_1(t)|L\rangle\langle L| + \varepsilon_0(t)|R\rangle\langle R|. \tag{2}$$

In (2), $k$ is the tunneling coupling coefficient between the dots, $\varepsilon_{L(R)}=2\hbar\Omega|a_{L(R)}|^4$ is the energy for each dot with $\Omega=e^2/(4\hbar C)$ corresponding to the angular frequency of one-half of the Coulomb charging energy of a quantum dot of capacitance $C$. When the quantum dot structure is exposed to an electrical field of the shape $E(t)=E_0 f(t)\cos(\omega t+\theta)$, where $E_0$ is the amplitude and $f(t)$ is the pulse envelope, the interaction between the dots is described by the time-dependent potential

$$V(t)=\frac{1}{2}\hbar\Omega_\omega f(t)\cos(\omega t+\phi)(|R\rangle\langle R|-|L\rangle\langle L|) \qquad (3)$$

where $\Omega_\omega=2eE_0 d/\hbar$ is the Rabi frequency, with $d$ being the centre-to-centre distance of the two dots. Therefore, in order to obtain the evolution of the complete system one has to solve the time-dependent Schrödinger equation

$$i\hbar\frac{d|\psi(t)\rangle}{dt}=[H_0(t)+V(t)]|\psi(t)\rangle. \qquad (4)$$

Substituting (1)-(3) in (4) and performing an appropriate unitary transformation, the following set of coupled differential equations is obtained [12],

$$\frac{da_L}{dt}=-ika_R+i\left[\frac{1}{2}\Omega_\omega f(t)\cos(\omega t+\theta)+\Omega(|a_R|^2-|a_L|^2)\right]a_L \qquad (5)$$

$$\frac{da_R}{dt}=-ika_L-i\left[\frac{1}{2}\Omega_\omega f(t)\cos(\omega t+\theta)+\Omega(|a_R|^2-|a_L|^2)\right]a_R \qquad (6)$$

The set of coupled equations (5) and (6) may be numerically solved by using the Runge-Kutta method, as it was done in [12]. Instead of solving (5)-(6) directly, we looked for a simpler set of coupled equations using the Lagrangian formalism. The Lagrangian used is,

$$L=-\left(a_L^*\frac{da_L}{dt}+a_R^*\frac{da_R}{dt}\right)-ik\left(a_L^*a_R+a_R^*a_L\right)-i\frac{\Omega}{2}\left(|a_L|^2-|a_R|^2\right)^2+iF(t)\left(|a_L|^2-|a_R|^2\right) \qquad (7)$$

$$F(t)=\frac{1}{2}\Omega_\omega f(t)\cos(\omega t+\varphi) \qquad (8)$$

One can easily check that (5) and (6) can be obtained from (7) using the Euler-Lagrange equations

$$\frac{\partial L}{\partial a^*_{L(R)}} - \frac{d}{dt}\frac{\partial L}{\partial\left(\frac{da^*_{L(R)}}{dt}\right)} = 0. \quad (9)$$

In order to obtain a simpler set of coupled equations, without ignoring its physical sense, we chose the following ansatz,

$$a_L(t) = \cos[\alpha(t)]e^{i\lambda(t)} \quad (10)$$
$$a_R(t) = \sin[\alpha(t)]e^{i\xi(t)} \quad (11)$$

By substituting these expressions in the proposed Lagrangian (7), the new Lagrangian obtained is

$$L = -i\left(\frac{d\lambda}{dt}\cos^2\alpha + \frac{d\xi}{dt}\sin^2\alpha\right) - ik\sin(2\alpha)\cos(\xi-\lambda) - i\frac{\Omega}{2}\cos^2(2\alpha) + iF(t)\cos(2\alpha). \quad (12)$$

Now, by putting (12) into the Euler-Lagrangian equations

$$\frac{\partial L}{\partial x} - \frac{d}{dt}\left(\frac{\partial L}{\partial\left(\frac{dx}{dt}\right)}\right) = 0, \quad (13)$$

where $x = \alpha$, $\lambda$ and $\xi$, one gets

$$\frac{d\phi}{dt} = -2F(t) - 2k\cos\phi\frac{1}{\tan(2\alpha)} + 2\Omega\cos(2\alpha) \quad (14)$$

$$\frac{d\alpha}{dt} = -k\sin\phi \quad (15)$$

where $\phi = \xi - \lambda$. Thus, (14)-(15) form the set of coupled differential equations that governs the dynamical behaviour of an electron between two quantum dots, when it is under the influence of an external electric field, represented by $F(t)$.

## 3. Numerical simulations

Following earlier studies [12], we start by considering the AC external field with a constant envelope function, $f(t)=1$. For this case, we have performed many simulations, varying the parameters, in order to see if our results were in agreement with those shown in [12]. Thus, it was initially neglected the Coulomb charging energy influence ($\Omega$ is set to zero) and considered an AC driven field with $\Omega_\omega/\omega = 5.05$, where $\Omega_\omega$ is the Rabi frequency and $\omega$ is the angular frequency of the field. Under these conditions, the system presents an oscillatory behaviour, as it is shown in Fig. 1. This picture shows the electron moving from one dot to the other, once the probabilities of being found in $Q_R$ and $Q_L$ quantum dots are alternating.

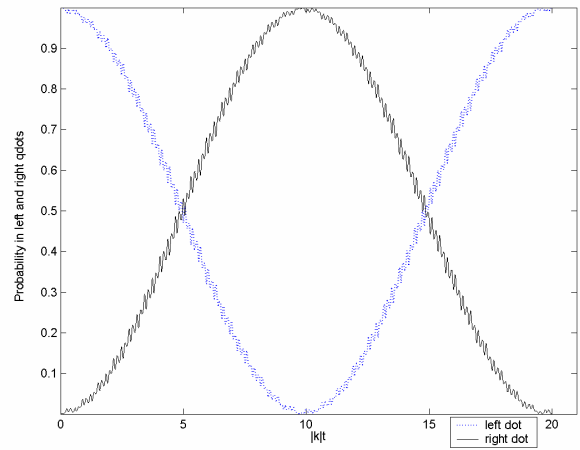

Fig. 1 – Probability of finding the electron in the $Q_L$ (initially occupied dot) and $Q_R$ dot versus time. Parameters for this simulation: $\Omega = 0$, $\Omega_\omega/\omega = 5.05$.

The second interesting case to check is the one which shows the destruction of tunneling, seen in Figure 2. In this figure, one clearly can see the electron remains at the same dot in which it was placed at the beginning. In fact, this is called electron localization. For this simulation, the parameters $\Omega$ and $\Omega_\omega/\omega$ were set to $1.9|k|$ and 5.05, respectively.

Changing the Rabi frequency to $0.9\omega$, the probability of finding the electron in each quantum dot assumes the profile seen in Figure 3, that is, the electron localization gets worse in this case.

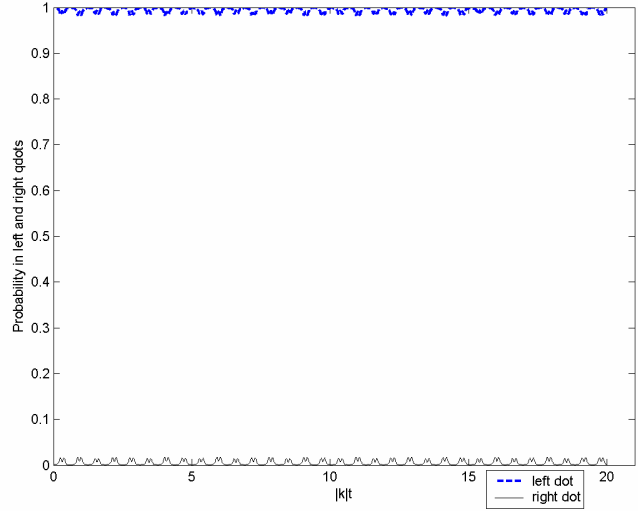

Fig. 2 – Probability of finding the electron in $Q_L$ and $Q_R$ dots *versus* time. One can see in this picture that the electron localization is determined. The parameters for this simulation are: $\Omega=1.9\cdot|k|$, $\Omega_\omega/\omega = 5.05$.

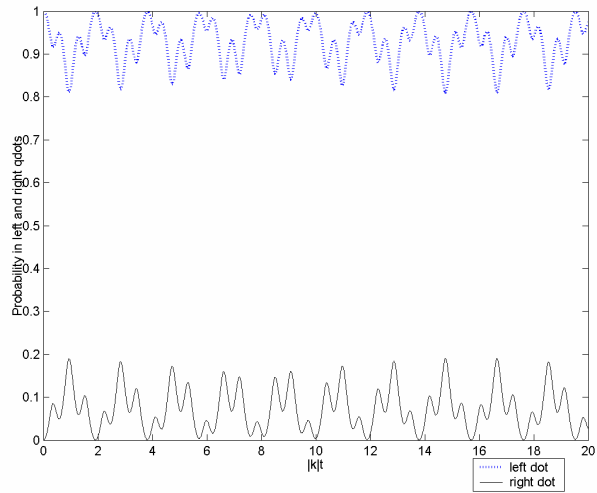

Fig. 3 – Probability of finding the electron in $Q_L$ and $Q_R$ dots *versus* time. One can see in this picture that the electron localization is also determined, but with lower certainty than that of Figure 2. The parameters for this simulation are: $\Omega = 0.9\cdot|k|$, $\Omega_\omega/\omega=5.05$.

Based on a simple comparison between Figures 2 and 3, one can conclude that the degree of electron localization is dependent on the set of parameters $\Omega$ and $\Omega_\omega/\omega$, for a given external AC driven field. If we consider that the degree of electron localization may be taken as (*1 - the variance of the probability curves*), for the set of parameters of Figure 2 the degree of electron localization is near 98%, while in Figure 3 it is approximately 81%. Similar curves are seen in figure 4, in which the Rabi frequency is kept in 5.05$\omega$ and the Coulomb charging energy assumes other two values, 0.17|k| and 0.6|k|. If one chooses to maintain the Coulomb charging energy in a fixed value, by changing the Rabi frequency electron localization and tunneling are still achievable. In this case (simulation not shown in the paper) what changes is the profile of the curve (if a tunneling is occurring) or the degree of electron localization (if destruction of tunneling takes place).

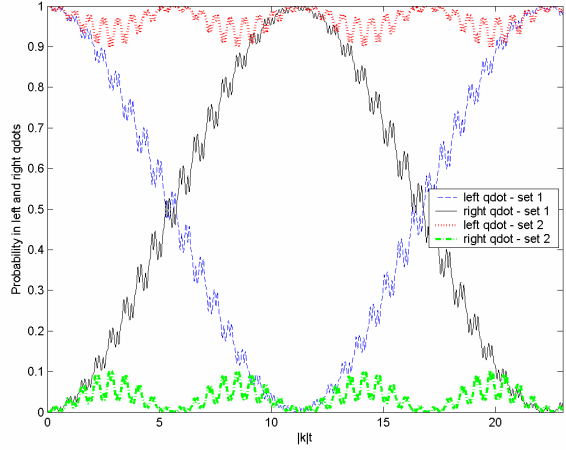

Fig. 4 – Probability of finding the electron in $Q_L$ and $Q_R$ dots *versus* time for two sets of parameters. For curves red and green the parameters are: $\Omega$=0.6·|k|, $\Omega_\omega/\omega$ =5.05, and for the other, $\Omega$=0.17·|k|, $\Omega_\omega/\omega$ = 5.05. Blue and red colors refer to the left dot.

All the cases shown up to this moment are related to free-tunnelling and/or electron localization, based only in the system parameters. Another interesting situation to observe is the possibility of control the electron localization, that is, transfer the electron from one dot the other and be sure that it will be there until another interaction with the system. To study this case, it is considered the semi-infinite pulsed AC field with an envelope function having a hyperbolic tangent shape, $f(t)$=tanh($t/\tau$), where $\tau$ is the parameter which controls the rise time of the pulse. The simulations using this kind of external excitation have their results presented in Figs. 5, 6 and 7. For the simulation in Fig. 5, the Coulomb charging energy is neglected, the Rabi frequency is equal to 2.4$\omega$ and $\tau$ is equal to 2/|k|.

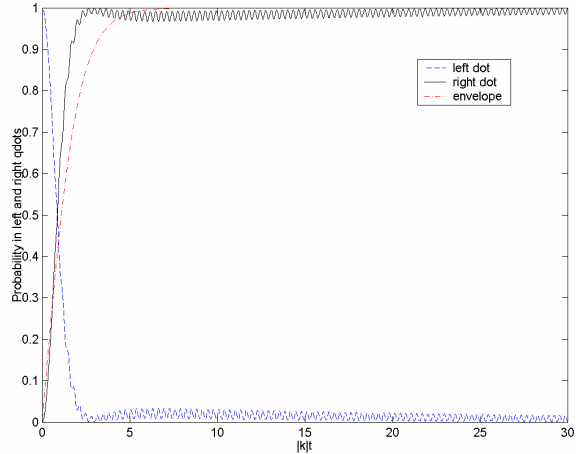

Fig. 5 – Probability of finding the electron in the left and right dot *versus* time for an envelope having hyperbolic tangent shape. For this simulation the parameters are: $\Omega = 0 \cdot |k|$, $\Omega_\omega/\omega = 2.4$ and $\tau = 2/|k|$.

As can be seen in Fig. 5, this is no more a tunnelling case and neither a simple case of confinement in the dot initially occupied, but this is in fact a transfer process in the sense that the electron changes from the $Q_R$ dot to the $Q_L$ dot and remains there. This is of great interest to quantum information processing, once it can be used as part of a controllable memory or a way to set the initial state of a quantum register. Figure 6 also shows electron transfer process, but for the simulation presented in this figure, the pulse arises slower, $\tau$ is equal to $5.2/|k|$. Thus, while the AC field has its intensity growing, the electron is free to tunnel, but when the envelope function reaches a value near the unity, the transfer occurs. From this figure one can conclude that the control over the system state is strongly related to the rise time of the pulse, and this gives a high-degree of control over the system to who manipulates the AC driving field.

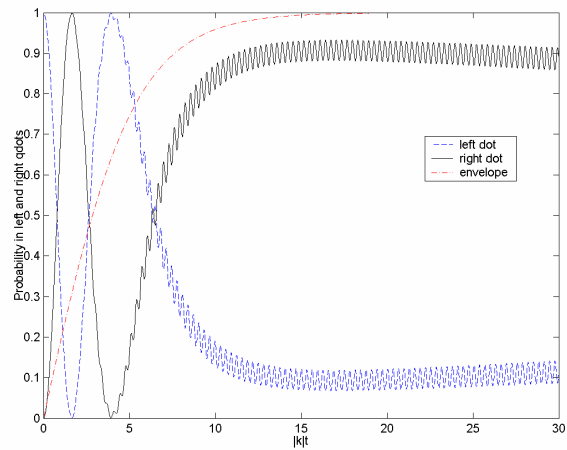

Fig. 6 – Probability of finding the electron in the left and right dot *versus* time for an envelope having hyperbolic tangent shape. For this simulation the parameters are: $\Omega = 0 \cdot |k|$, $\Omega_\omega/\omega = 2.4$ and $\tau = 5.2/|k|$.

To finish the electron transfer process, in Fig. 7 the Coulomb charging energy is considered, $\Omega=0.675|k|$, the Rabi frequency is equal to $5.52\omega$ and $\tau$ is equal to $5.2/|k|$. The main result in this figure is that the electron localization in the final quantum dot gets worse when Coulomb charging energy is taken into account.

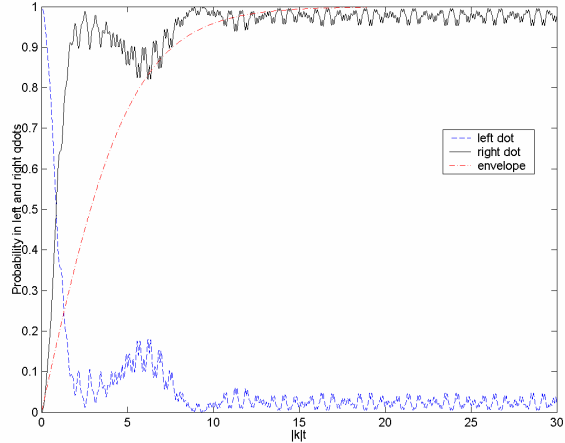

Fig. 7 – Probability of finding the electron in the left and right dot *versus* time for an envelope having hyperbolic tangent shape. For this simulation the parameters are: $\Omega = 0.675 \cdot |k|$, $\Omega_\omega/\omega = 5.52$ and $\tau = 5.2/|k|$.

## 4. Application of the control of electron localization in quantum-dot cellular automata digital circuits

Quantum dot cellular automata, QCA, has been proposed for building of digital logic circuits at nanoscale [13-14] and it is one of the promising technologies for solid-state quantum computing implementation [15]. The basic block of a QCA is a cell consisting of four quantum dots – a pair of coupled quantum dot structures – having two extra electrons, one for each pair of coupled quantum dots. In a pair of coupled quantum dots, the electron can tunnel from the top dot to the bottom dot or vice-versa, but it is not allowed one electron to tunnel from one pair to the other. Due to the Coulomb repulsion between the extra electrons, the lowest energy configurations are those where the electrons occupy the opposite sites of a diagonal. These situations can be regarded as classical bits '0' and '1' and they represent a basis for the quantum state of the cell. Traditionally, the state of a QCA cell has been represented by its polarization, *P*, and the two classical states are $P=+1$ (bit 0) and $P=-1$ (bit 1). If the cell is in the quantum state $\alpha|0\rangle+\beta|1\rangle$, then its polarizations is simply given by $P=\langle\sigma_z\rangle=|\alpha|^2-|\beta|^2$. Some configurations of a QCA cell are presented in Fig. 8.

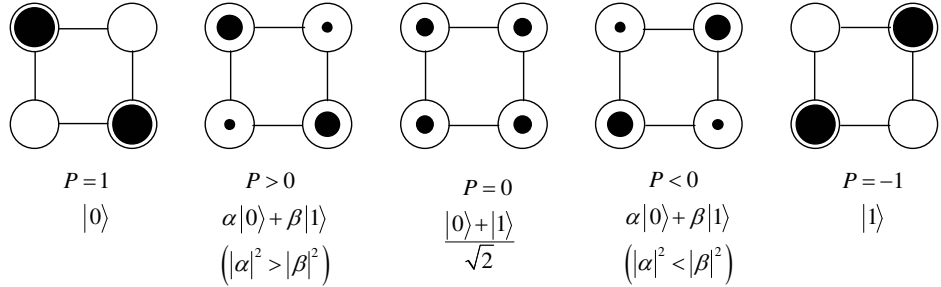

Fig. 8 – Some quantum cellular automata configurations.

The dynamics of a single QCA cell and a three qubit controlled operation using three QCAs were described in [15].

In all those works about QCA, the QCA cell is supposed to be perfectly assembled such that, the moving of an electron in a coupled dot structure will always effect the electron at the other coupled dot structure in such way that, if one waits enough time, the QCA cell will always move to one of the ground states. Here we are going to consider a QCA where only the left coupled quantum dots structure has its tunneling energy externally controlled by an electrical field and the tunneling energy of the right coupled quantum dots structure is constant in a value that makes the probability of tunneling equal to ½. In order to model the switching behavior of this QCA cell, we use the following quantum state,

$$|\Psi(\eta)\rangle = \left[\frac{1}{2}\exp\left(-\frac{\tau\eta}{1-\eta}\right)\right]^{1/2} a_R|TT\rangle + \left[1-\frac{1}{2}\exp\left(-\frac{\tau\eta}{1-\eta}\right)\right]^{1/2} a_R|TB\rangle$$
$$+ \left[1-\frac{1}{2}\exp\left(-\frac{\tau\eta}{1-\eta}\right)\right]^{1/2} a_L|BT\rangle + \left[\frac{1}{2}\exp\left(-\frac{\tau\eta}{1-\eta}\right)\right]^{1/2} a_L|BB\rangle \quad (16)$$

where, now, instead of R and L we are using T (top) and B (bottom) since the coupled quantum dots are in the vertical. In (16), $\eta$ (0≤$\eta$≤1) is the parameter that takes into account the coupling between the double dots structures that form the QCA, while $\tau$ is the constant of decayment whose value can be determined experimentally. Basically, $\tau$ says how long one has to wait before the QCA cell reaches a ground state. The limit situations are decoupled pairs, $\eta=0$, and maximally coupled pairs, $\eta=1$, resulting, respectively, in the following states:

$$|\Psi(\eta=0)\rangle = \left(a_T|T\rangle + a_B|B\rangle\right) \otimes \left(\frac{|T\rangle+|B\rangle}{\sqrt{2}}\right). \quad (17)$$

$$|\Psi(\eta=1)\rangle = \left(a_T|TB\rangle + a_B|BT\rangle\right) \quad (18)$$

The state in (18) is the one that is always taking into consideration ($|TB\rangle$ bit 0 and $|BT\rangle$ bit 1) in QCA literature. Observing (16) one can still define the polarization as

$$P = \left(|a_T|^2 - |a_B|^2\right)\left[1 - \exp\left(-\frac{\tau\eta}{1-\eta}\right)\right]. \quad (19)$$

Once the QCA was assembled and the parameter $\eta$ was defined, one can use (16) to model its quantum state. In particular, for $\eta=1$, the state (18) shows us that we can use the electron localization described in Section 3 to set up the value of the QCA, with $|a_T|^2$ and $|a_B|^2$ described by equations (10)-(15). This allows us to construct configurable digital logic circuits. An example is presented in Fig. 9 where the QCA array can be configured to be a AND gate ($c=0$) or an OR gate ($c=1$), depending on the value of the QCA of control [16].

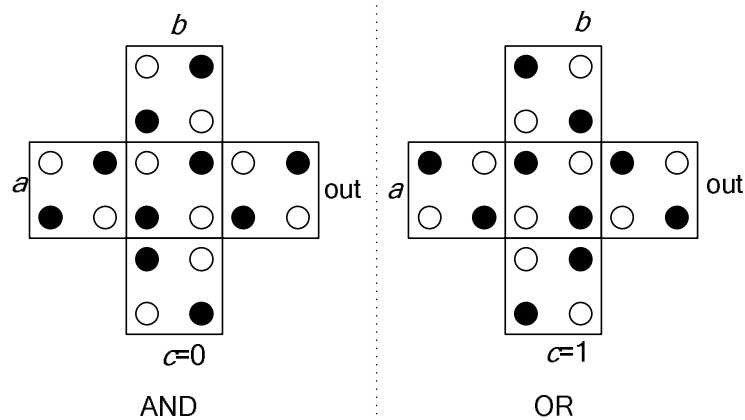

Fig. 9 – QCA-made configurable digital logic circuit through electron localization.

Hence, controlling the electron localization of the left coupled double quantum-dot structure of the QCA of control, permits one to select between AND and OR functions.

## 5. Conclusions

We have proposed a variational formulation in order to solve numerically the set of equations that governs the dynamic of an electron in a double quantum dot structure under influence of an AC electric field. The results obtained are consistent with those found in the literature. Furthermore, although it had not been shown in the text, the solution of the set of equations (14)-(15) requires lower computational effort than a direct solution of equations (5)-(6) using Runge-Kutta numerical method. The results of the simulations showed clearly that electron localization is possible. This is very important for realization of classical and quantum computing with quantum dots. In particular, we considered the dynamic of quantum cellular automata formed by a pair of coupled quantum dots structures through a phenomenological parameter $\eta$, that models the interaction between both coupled quantum dot structures. When $\eta=1$ the interaction is maximum and the electron localization in one coupled quantum dot structure controls the bit value that the whole QCA represents. This permits one to construct configurable digital logic using QCA and electron localization.

## References

[1] C.E. Creffield and G. Platero, Phys. Rev. B 69, 165312, 2004.

[2] A. Kiraz, M. Atatüre and A. Imamoglu, Phys. Rev. A 69, 032305, 2004.

[3] C. Santori, D. Fattal, J. Vuckovic, G. S. Solomon and Y. Yamamoto, New J. of Phys 6, 89, 2004.

[4] J. Vuckovic, C. Santori, D. Fattal, M. Pelton, G. S. Solomon, B. Zhang, J. Plant and Y. Yamamoto, Electron Devices Meeting, 2002. IEDM'02, pp. 87-90, 8-11 december, 2002.

[5] T. M. Stace, G. J. Milburn and C. H. W. Barnes, Phys. Rev. B 67, 085317, 2003.

[6] R.M. Stevenson, R.M. Thompson, A.J. Shields, I. Farrer, B.E. Kardynal, D.A. Ritchie and M. Pepper, Physical Review B 66, 081302(R), 2002.

[7] I. Marcikic, H. de Riedmatten, W. Tittel, H. Zbinden, and N. Gisin, Nature 421, 509 – 513, 2003.

[8] D. Fattal, E. Diamanti, K. Inoue and Y. Yamamoto, Phys. Rev. Lett. 92, 037904, 2004.

[9] R.L. de Visser and M. Blaauboer, Phys. Rev. Lett. 96, 246801, 2006.

[10] D. Loss, D. P. DiVincenzo, Phys. Rev. A, 57, 1, 120, 1998.

[11] I. D'Amico, e-print arXiv:cond-mat/0511470, available at [www.lanl.gov](www.lanl.gov), 2005.

[12] E. Voutsinas, A.F. Terzis and E. Paspalakis, J. of Mod. Opt., 51, n° 4, 479, 2004.

[13] K. Walus and G. A. Jullien, Proc. of the IEEE, 94, 6, 125, 2006.

[14] C. S. Lent and P. D. Tougaw, Proc. of the IEEE, 85, 4, 541, 1997.

[15] G. Tóth and C. S. Lent, Phys. Rev. A, 63, 052315, 2001.

[16] W. J. Townsend and J. A. Abraham, 4th IEEE Conference on Nanotechnology, pp. 625-627, Munich, Germany, August 17-19, 2004.